\documentclass[preprint,pre,aps,superscriptaddress,a4paper]{revtex4}
\usepackage[english]{babel}
\usepackage[utf8]{inputenc}
\usepackage{graphicx}
\usepackage{amsmath}
\pdfoutput=1

\begin{document}
\title{Triangular lattice models for pattern formation by core-shell particles with different shell thicknesses}
\author{V.S.Grishina}
    \email[Correspondence email address: ]{vera1grishina@gmail.com}
    \affiliation{Belarusian State Technological University, 13a Sverdlova str., 220006 Minsk, Belarus}
\author{V.S.Vikhrenko}
    \email[email address: ]{vvikhre@gmail.com}
    \affiliation{Belarusian State Technological University, 13a Sverdlova str., 220006 Minsk, Belarus}
\author{A.Ciach}
    \email[email address: ]{aciach@ichf.edu.pl}
    \affiliation{Institute of Physical Chemistry,
	Polish Academy of Sciences, Kasprzaka 44/52, 01-224 Warszawa, Poland}
\date{\today} 

\begin{abstract}
Triangular lattice models for pattern formation by hard-core soft-shell particles at interfaces are introduced and studied
in order to determine the effect of the shell thickness and structure. In model I,
we consider particles with  hard-cores covered by  shells of cross-linked polymeric chains. 
In model II, such inner shell is covered by a 
much softer outer shell. 
In both models,  the hard cores can occupy sites of the triangular lattice, and nearest-neighbor repulsion following from
overlaping shells is assumed.
 The capillary force is represented by
the second- or the fifth neighbor attraction in model I or II, respectively.
Ground states with  fixed chemical potential $\mu$ or with fixed fraction of occupied sites $c$
are thoroughly studied.
For $T>0$, the $\mu(c)$ isotherms, 
compressibility and specific heat are calculated by Monte Carlo simulations. 
In model II, 6 ordered periodic patterns occur in addition to 4 phases found in model I. 
These additional phases, however, are stable only 
at the phase coexistence lines, i.e. in regions of zero measure at the $(\mu,T)$ diagram which otherwise looks like the diagram
of  model I.  In the canonical ensemble,
these 6 phases and interfaces between them appear in model II for large intervals of $c$, and the number of possible patterns is much larger 
than in model I.
We calculated surface tensions for different interfaces, and 
found that the 
favourable orientation of the  interface corresponds to its
smoothest shape in both models. 
\end{abstract}


\maketitle
\section{Introduction}

The statistical-mechanical theory of molecular systems based on interaction potentials of a Lenard-Jones type was 
developed in the second half of the last century \cite{Perkus,Weeks,Barker,Hansen,Evans,Ben}.
Later, attention of scientists was shifted to more complex systems \cite{Israel, Mezzenga} including solutions of 
colloids, proteins, etc., in particular with interparticle interactions described by the potentials of DLVO 
(Derjagin-Landau-Verwey-Overbeek) type \cite{Der,Verwey}. Recently, complex fluids containing nanoparticles of nontrivial
shape, structure and chemistry are intensively studied, because they  
show fascinating properties,
and can find numerous practical applications.

In particular, colloid metal or semiconducting nanoparticles find numerous applications in  catalysis, optics, biomedicine, 
environmental science, 
smart materials, etc. In these applications, it is important to prevent the nanoparticles from aggregation.
In addition to traditional charge-stabilized colloids,  recently another method of keeping desired separation between the particles is
becoming popular. Namely, various types of core-shell particles are produced~
\cite{vasudevan:18:0,isa:17:0}.
In the core-shell particles,
the surface of the hard, typically metal, magnetic or silica nanoparticle is covered by a polymeric shell. 
The shell can be organic or inorganic, which strongly influences interaction of the monomers with the solvent particles.
The interactions with the solvent and the entropic  effects can lead to a shrunk or to a swollen shell 
for high or low temperature. Such behavior was observed for Au@pNIPAM  core–shell particles~\cite{contreras:09:0}. 
The hybrid core-shell particles can have
both the cores and the shells of different and controlled sizes, and the softness of the shells can be controlled 
in particular by the crosslinking
of the polymeric chains. The effective interactions between the core-shell particles strongly depend on the thickness,
architecture  and chemistry of the shells. These properties can be tuned and adjusted to the desired effective interactions.

Particularly important are monolayers of the core-shell particles on various 
interfaces\cite{vasudevan:18:0,isa:17:0,vogel:12:0,nazli:13:0,volk:15:0,honold:15:0,geisel:15:0,rauh:16:0,karg:16:0}. 
They can find applications in plasmonic systems, anti-reflecting coating, pre-patterned substrates 
for growing ordered structures or for sensing.
At the interface of two liquids, the polymeric shell becomes deformed. In addition, capillary forces 
mediated by the interface 
appear. In the case of particles with hydrophilic shells on oil-water interface, 
the polymeric chains tend to be in contact with the aqueous 
rather than with the oil 
phase, and the core-shell particles look like a fried egg\cite{rauh:16:0,isa:17:0}. 

Experiments usually show  hexagonal arrangement of the particles at the interface, with the distance between the
particles equal or larger than their diameter. Growing pressure or density can lead to isostructural transition 
to  the hexagonal phase with a smaller unit cell 
\cite{rey:16:0,isa:17:0}. When
the density increases, the  distance between the particles can become smaller than the diameter of the particle, 
because the chains
surrounding different particles can interpenetrate, and the
shells can be deformed.
The smallest possible distance between the particles is equal to the hard-core diameter. 
In some cases, however, different patterns, including clusters or voids, are formed for intermediate density~
\cite{vasudevan:18:0,rauh:16:0,rey:16:0,isa:17:0}. 
The presence of empty regions
surrounding the hexagonal arrangements of particles observed in some experiments~\cite{rauh:16:0},
suggest that the effective potential
between the particles takes a minimum for a certain distance between them.

 Fully atomistic modeling of the pattern formation by the core-shell particles is very difficult, 
 but possible~\cite{camerin:19:0}. Unfortunately, the atomistic modeling is
   restricted to a particular example of the density of chains attached to the metallic core, chain length, chemistry
and architecture. 
Because very large number of different types of shells is possible in experiment, there is a need for a simplified,
coarse-grained theory that could predict general trends in pattern formation for various  ranges,
strengths and shapes of the effective potential.
The particles can move freely in the interface area, but out-of-plane mobility is reduced. For this reason,
the particles trapped at the interface can be modeled as a two-dimensional system.

Lattice models allow for much simpler calculations and faster simulations, and it is much easier to gain a general overview 
by considering a class of lattice models for core-shell particles at interfaces.
The study of lattice models for hard-core soft-shell particles at interfaces 
was initiated in Ref.\cite{ciach:17:0}. In the model considered in Ref.\cite{ciach:17:0}, 
the lattice constant is equal to the hard-core diameter, and the multiple occupancy of lattice sites is forbidden.
Next, a soft repulsion
at small distances (nearest-neighbors) is followed by an attraction  at larger distances (second or third neighbors). 
Exact results for the one-dimensional lattice
model show good agreement of the density-pressure isotherms with experiments of Ref.\cite{rauh:16:0}. 
At the same time, strong dependence of the formed structures on the range of attraction is observed. 
A 2D system was next modeled on a triangular lattice, with nearest-neighbor repulsion and third-neighbor
attraction\cite{groda:19:0}. Earlier, the criticality of the system with the nearest neighbor repulsion and the
second neighbor attraction on a triangular lattice was investigated in Refs.~\cite{mihurat:77:0,landau:83:0}, motivated
by experimental results concerning gas adsorption on graphite as well as
 by a purely theoretical interest. 

In this work we study in detail  triangular lattice models with nearest-neighbor repulsion, and second-neighbor 
or fifth-neighbor attraction. In the first model, the  shell is relatively thin. In the second model, the  inner shell
is harder than the very soft outer shell, and the particles are larger than in the first model. The models are introduced in sec.\ref{model}. In the same section we discuss for what properties
of the core-shell particles their distribution on an interface can be described by our models at least on a qualitative level. 
In sec.\ref{GSIa} and \ref{GSIIa} we study  the ground state (GS), i.e. we
determine the ordered phases in the two models for open systems at $T=0$. 
In sec.\ref{STI} and \ref{STII} we consider the GS with fixed number of particles 
for model I and II respectively, and
calculate the surface tension at different interfaces  at $T=0$. In sec.\ref{T>0}, we determine the concentration-chemical potential isotherms, 
specific heat, thermodynamic parameter (inverse compressibility), and order parameter (OP) for $T>0$.
The last section contains discussion and conclusions.

\section{The model}
\label{model}
\begin{figure}
        \centering
        \begin{minipage}{0.30\linewidth}
        	\includegraphics[scale=0.26]{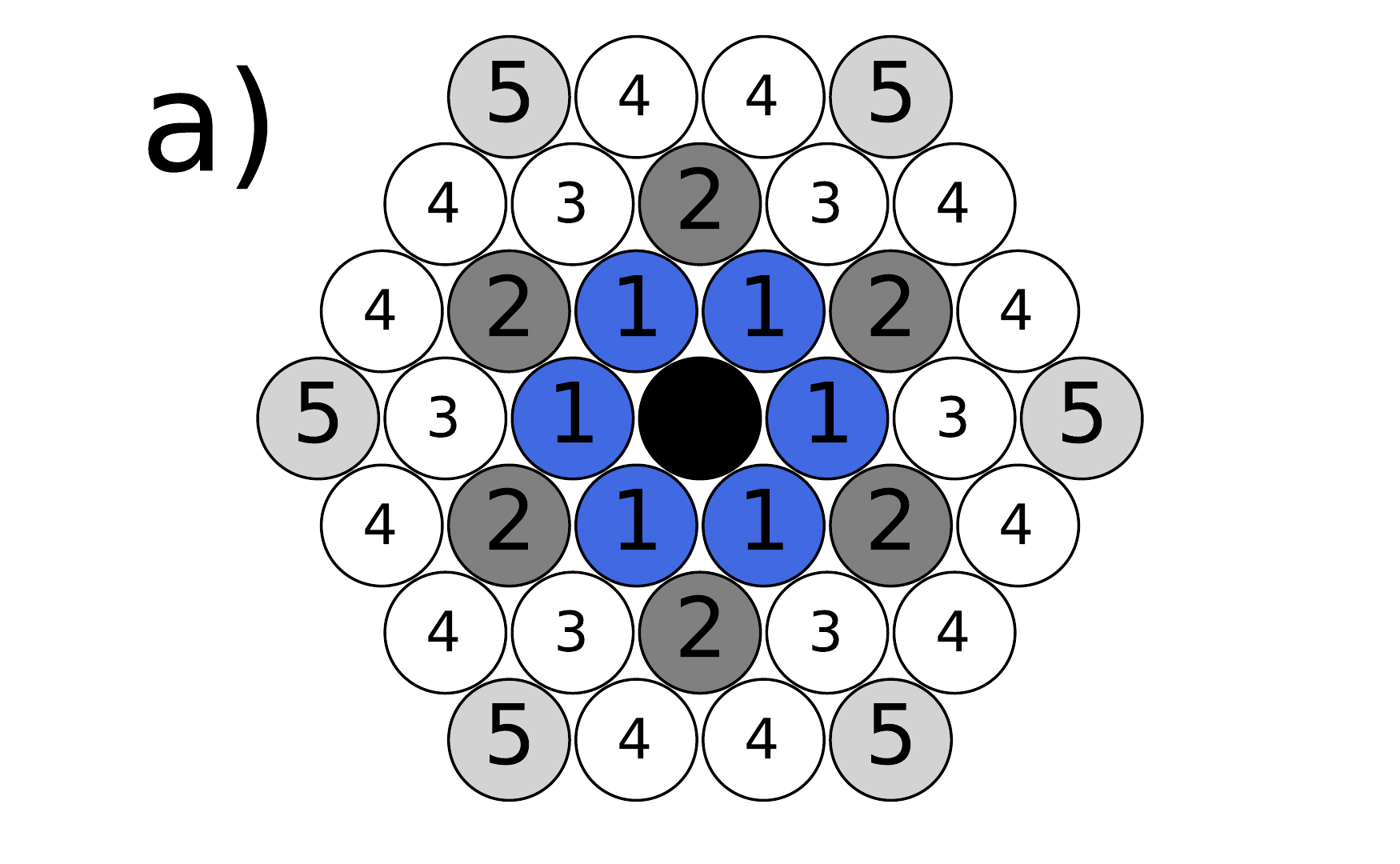}
        	\tiny \centering        	
        \end{minipage}
   		 \begin{minipage}{0.60\linewidth}
    	\includegraphics[scale=0.26]{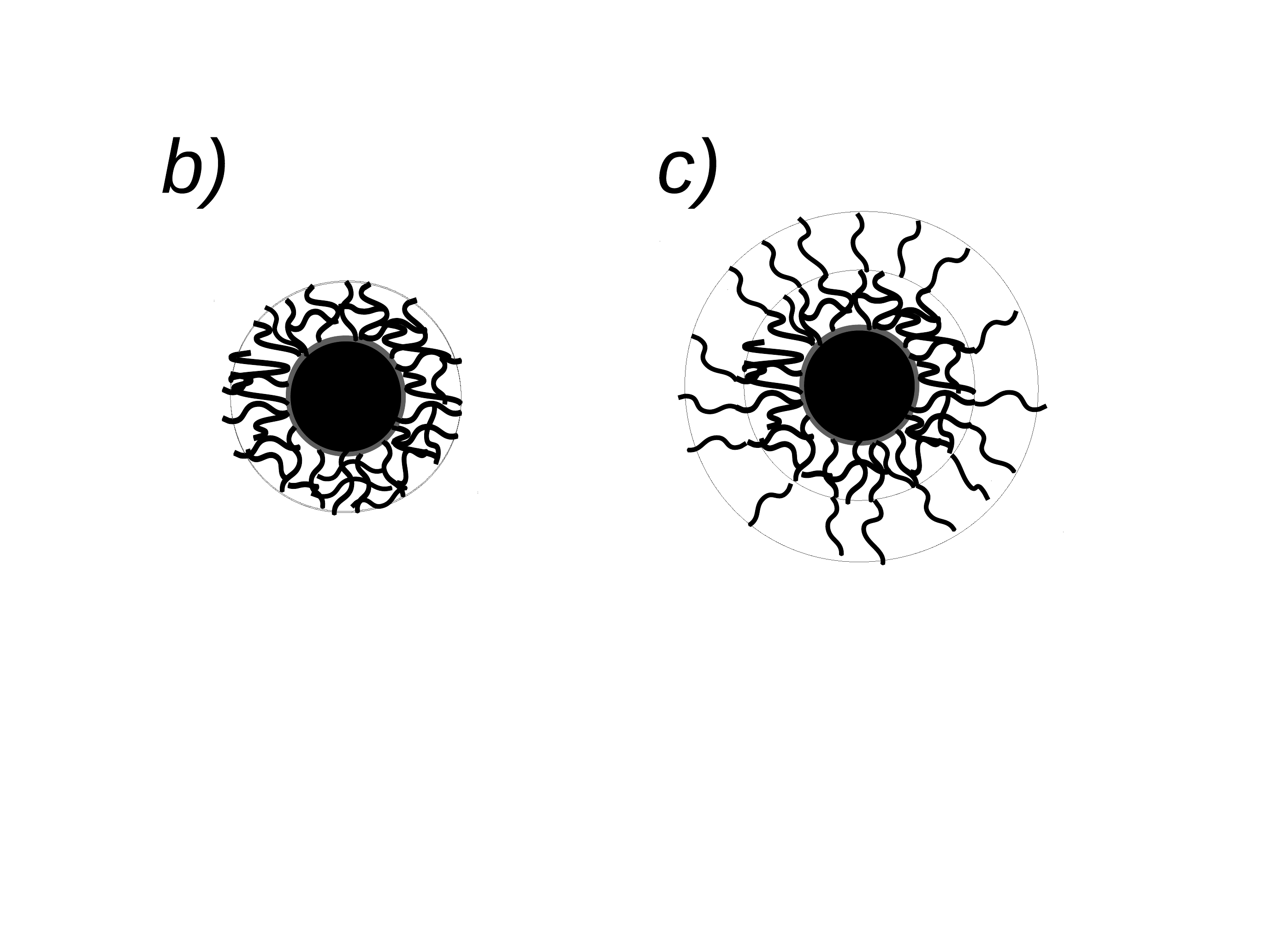}
    	\tiny 
       	\end{minipage}    
        \caption{(a) The first five coordination spheres around the central hard-core of the core-shell particle (black).
        (b) a cartoon showing schematically the core-shell particle in model I. (c) a cartoon showing schematically the core-shell particle in model II.}
        \label{fig:neighbors}
    \end{figure}

The model under consideration is a lattice fluid with particles occupying  sites of a triangular lattice containing $M$
lattice sites. The lattice parameter $a$ is equal to the diameter of the hard core of the particles. Multiple occupancy of the lattice sites is forbidden, since the cores cannot overlap. Particles that occupy 
the lattice sites on mutual coordination spheres of different order interact with each other.
The thermodynamic Hamiltonian of the open system with the chemical potential $\mu^*$ is: 
    \begin{equation}
    \label{H}
        H = \frac{1}{2} \sum_{k=1}^{k_{max}}\sum_{k_i=1}^{z_k}\sum_{i=1}^{M}J^*_k \hat{n_i}\hat{n_{k_i}}-\mu^*\sum_{i=1}^{M}\hat{n_i},
    \end{equation}
where $M$ is the total number of the lattice sites, $k_i$ enumerates the sites of the $k$-th coordination sphere around the site $i$,
$z_k$ is the coordination number, $J^*_k$ is the interaction constant for the $k$-th coordination sphere, $\hat{n_i}$ is 
the occupation number (0 or 1), and $\mu^*$ is the chemical potential. 
The first five coordination spheres are shown in Fig.\ref{fig:neighbors}a.
    
We consider the simplest case of competing interactions with nearest-neighbor repulsion ($J^*_1=J$) and next nearest 
($J^*_2=-J_2J$) or fifth ($J^*_5=-J_5J$) neighbors attraction. 
The first case (model I) is appropriate for particles with thin, relatively stiff shells (Fig.\ref{fig:neighbors}b). 
The second case (model II), corresponds to particles that in addition to the inner shell have a much softer outer shell 
(Fig.\ref{fig:neighbors}c). 

As an energy unit we choose the strength of repulsion, $J$, and introduce the dimensionless chemical potential by $\mu=\mu^*/J$.
By $J_a$ we shall denote the strength of the attraction in $J$ units,
i.e. $J_a=J_2$ or $J_a=J_5$ for model I or II, respectively.

The nearest-neighbor repulsion is suitable for a coarse-grained model of particles that have cross-linked polymeric shells of
a thickness comparable with the radius of the hard core of the particle.
The  shells can overlap and be deformed at some energetic cost, equal to $J_1^*$ when the hard-cores of two particles are in contact. We assume such shells for both models.

In the first model, the particle consists of the hard core and of the above described shell. 
The attraction between the second-neighbors follows from the effective capillary forces.
We assume that the effective potential takes the  minimum when the shells of the particles are in contact.
In model I this is the case when the second neighbors on the lattice are occupied.

In the second model, the inner shell of cross-linked chains is followed by a much softer outer shell consisting of relatively 
few polymeric chains. The chains attached to different particles
can interpenetrate at much lower energetic cost than the chains of the inner shells. 
The  sum of all effective interactions  between the particles at the corresponding distances
(second, third and fourth neighbors on the lattice)  can be neglected. Finally, attraction between the fifth neighbors 
follows from the capillary forces. 
  
In Fig.\ref{fig:neighbors}, the black central circle represents the core of the particle. In the first model, the radius of
the core-shell particle is $\sqrt 3/2$ (in units of $a$). In the second model, the radius of
the core-shell particle is $3/2$. 

In the next two sections we consider models I and II at zero temperature and determine the  GS first
for an open system, and next for fixed number of particles. The equilibrium structures correspond to
the minima of the thermodynamic Hamiltonian defined in Eq. (\ref{H}) divided by the number of the lattice sites. 
We denote the  thermodynamic Hamiltonian per lattice site in units of $J$ by $\omega=H/(JM)$.
We consider the two models separately, starting from the first, simpler model.

\section{The GS of model I (thin shells)}
\label{GSI}

\subsection{The GS of an open system}
\label{GSIa}

\begin{figure}
        \centering
        \includegraphics[scale=0.8]{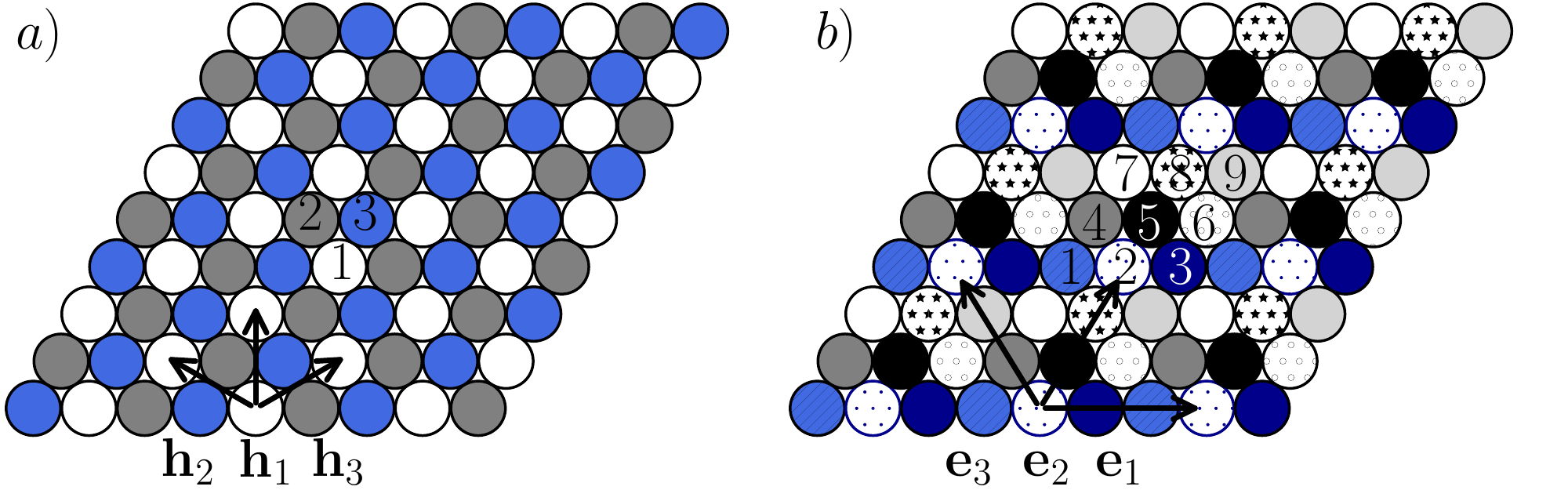}
        \caption{
        (a) the 3 sublattices of model I and their lattice vectors ${\bf h}_i$ for $i=1,2,3$. 
        (b) the 9 sublattices of model II and their lattice vectors ${\bf e}_i$ for $i=1,2,3$. 
     The sites belonging to the $n$-th sublattice in the unit cell are labelled $n$.
     Note that ${\bf e}_i\perp {\bf h}_i$.}
                      \label{fig:sublattices}
    \end{figure}

The system with the interactions up to the second neighbors can be split into three sublattices
(see Fig.\ref{fig:sublattices}a), and 
the allowed concentrations $c$ (the fraction of the sites occupied by the cores) of the GS are 0 
(vacuum), 1/3 (one of the sublattices is filled), 2/3 (two filled sublattices) and 1 (all sublattices are filled).

At the vacuum state $\omega(0)=0$. 
At the concentration $c=1/3$, the core of the particle has six next nearest neighbors and $\omega(1/3)=-J_2-\mu/3$
because a third part of the lattice sites is occupied, and each interaction bond is taken into account twice when calculating 
the total energy of the system. For $c=2/3$, $\omega(2/3)=1-2J_2-(2/3)\mu$ because each particle core has three nearest and six
next nearest neighbors. Finally, $\omega(1)=3-3J_2-\mu$ for the dense system.

\begin{figure}
	\centering
	\includegraphics[scale=0.4]{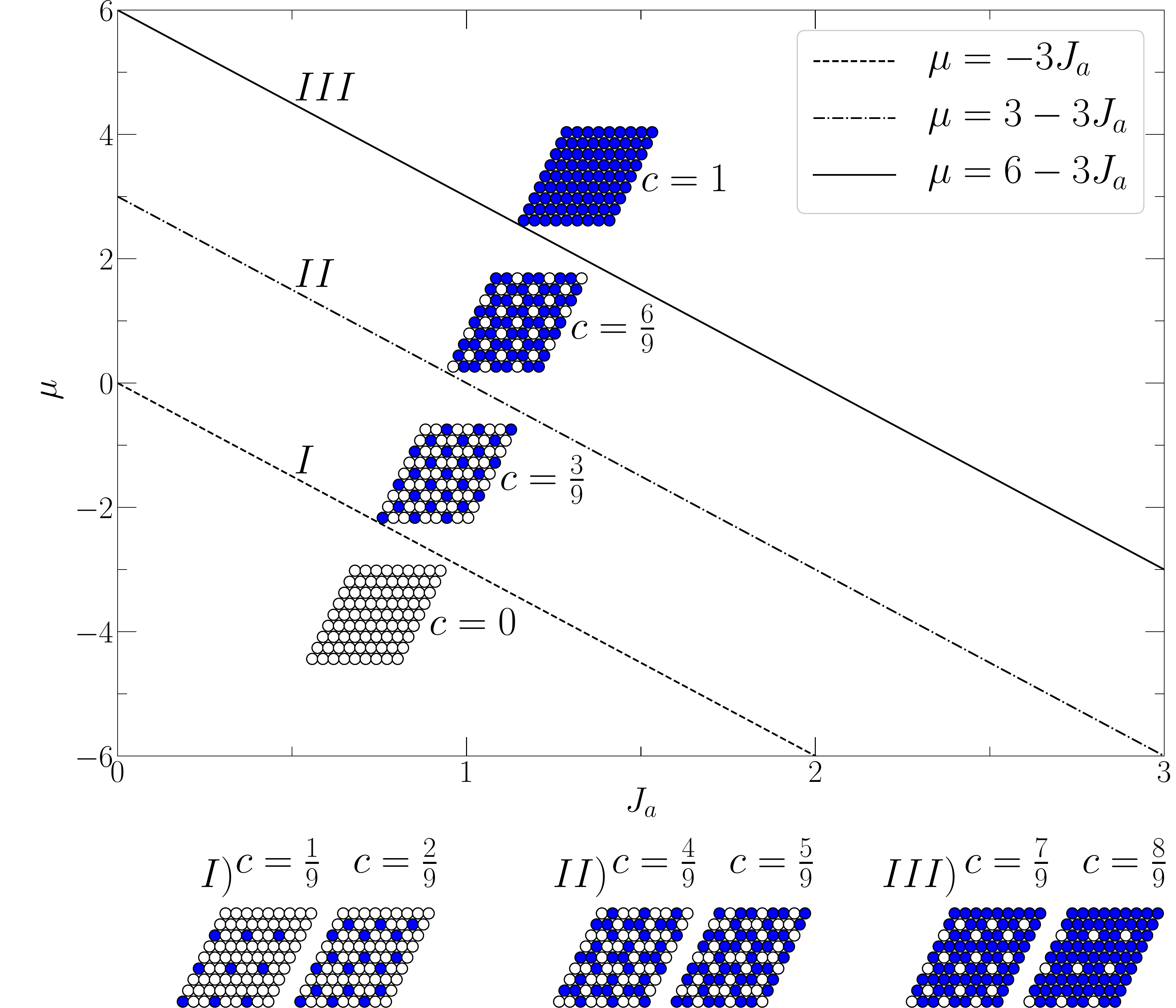}
	\caption{The ground states for models I and II, with the filled circles representing the hard-cores of the particles. 
	The structures shown in the insets are stable in the regions separated by the coexistence lines, as well as at these lines.
	For model I, $J_a=J_2$.
	For model II, $J_a=J_5$, and in addition to the phases shown in the diagram, 
	the phases with  $c=1/9$ and $c=2/9$ are stable
	along the dashed line labelled I, the phases with $c=4/9$ and $c=5/9$  are stable along the dash-dotted line labelled II,
	and the phases with $c=7/9$ and $c=8/9$ are stable at the line III (solid).
	The structure of the phases stable along the coexistence lines I, II, and III is shown 
	at the cartoons below the $(J_a,\mu)$ diagram.}
	\label{fig:GS}
\end{figure}

By comparing the above expressions for $\omega(c)$, we have found that the vacuum state
is stable for  $\mu\le-3J_2$, one of the sublattices is filled for $-3J_2\le\mu\le 3-3J_2$, two sublattices are filled for
$3-3J_2\le\mu\le 6-3J_2$, and the dense state exists for $\mu\ge 6-3J_2$. The phase diagram is shown in Fig.\ref{fig:GS}, 
and the structure of the ordered phases is shown in the insets.

Because  the ordered phases shown in Fig.\ref{fig:GS} are uniquelly chracterized by the concentration at $T=0$, 
the phase with the concentration $n/3$ 
will be refered to as ``the $c=n/3$ phase'', with a similar rule for model II, where ``the $c=n/9$ phase'' will
denote the ordered phase in which $c=n/9$ at $T=0$.

\subsection{The GS for fixed number of particles and the line tensions.}
\label{STI}
When in the system with periodic boundary conditions
the fixed number of particles $N$ is different from $M/3, 2M/3$ or $M$, 
then an interface between two coexisting phases must occur.
For comparable, macroscopic areas of the coexisting phases, the interface should have a form of a straight line.
The orientation of the spontaneously appearing interface  is determined by the
minimum of the  energy of the whole system, because at $T=0$ 
 the entropy plays no role. On the triangular lattice the distinguished orientations are parallel either
 to the vectors ${\bf e}_i$ or to the vectors ${\bf h}_i$ (see Fig.\ref{fig:sublattices}).

\begin{figure}
	\centering
	\includegraphics[scale=0.45]{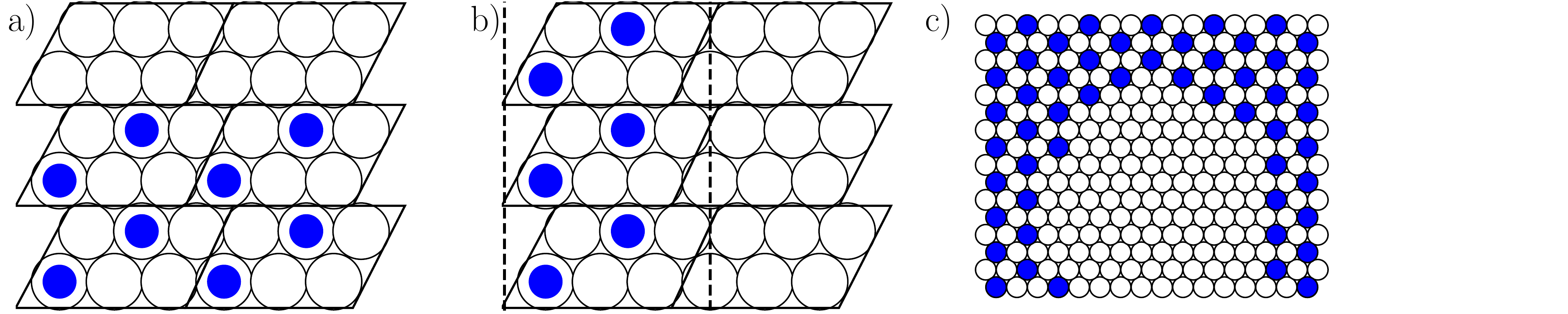}
	\caption{Panels (a) and (b):  
	The lattice with periodic boundary conditions in both, the vertical and the horizontal direction. 
	The interfaces between the vacuum and the $c=1/3$ phases (dashed lines) are parallel to ${\bf e}_1$ in (a) or to  ${\bf h}_1$ in (b). 
	The particles at the interface are connected by a zig-zag or by a straight  line in (a) or (b), respctively.
	Panel (c): A small part of a simulation snapshot of  model I for the fixed concentration $c=395/1296 \approx 0.3048<1/3$ 
	on the lattice 
	$36\times 36$ after 200 MCS annealing and 7000 MCS at $T=0.1$. The hole of 37 empty sites has almost ideal hexagonal
	shape
	with the sides parallel
	to the lattice vectors ${\bf h}_i$ (the image due to the periodic boundary conditions is not shown). 
	The total energy of the system is $-1163J_2$ 
	as compared to $-1164J_2$ for the ideal configuration at $T=0$.} 
	\label{unit_cells}
\end{figure}
\begin{figure}
	\centering
	\includegraphics[scale=0.7]{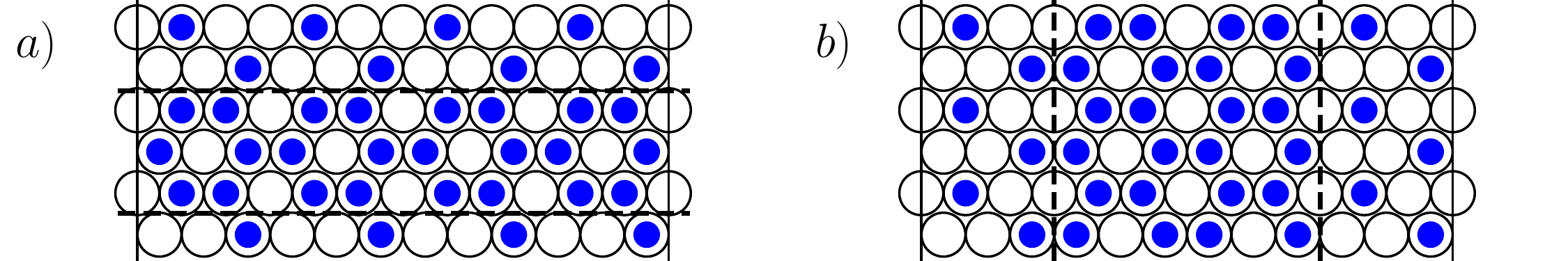}
	\caption{The interface lines (dashed) between the $c=1/3$ and $c= 2/3$ phases on a lattice
	with PBC conditions in directions
	${\bf e}_1$ (horizontal) and in direction perpendicular to ${\bf e}_1$ (vertical). 
	The periodic boundary conditions are fulfilled at the left 
	and right vertical lines that indicate the identified columns of sites.
	(a) The two interfaces are parallel to ${\bf e}_1$ (horizontal lines). 
	(b)  The two interfaces are perpendicular to ${\bf e}_1$ (vertical lines). 
	}
	\label{interphase}
\end{figure}

In order to calculate the line tension between two coexisting phases, we consider periodic boundary conditions in directions 
parallel and perpendicular to the interface, and allow for an integer number of unit cells of the periodic phase (or phases) in the two directions.  The interfaces parallel and perpendicular to ${\bf e}_1$ are shown in Fig.
\ref{unit_cells} and \ref{interphase}. 
Because of the periodic boundary conditions in the direction perpendicular to the interface, two parallel interfaces are formed.
 The dimensionless grand potential in the presence of the two interfaces takes the form
\begin{equation}
 \Omega=\omega M+2\sigma L,
 \label{sig}
\end{equation}
 where $\omega$ is the grand potential 
per lattice site in the coexisting phases in the bulk, $L$ is the length of the interface in units of $a$
and $\sigma$ is the dimensionless surface tension. 

Let us first consider the interface between the vacuum and the hexagonal phase with $c=1/3$. 
The interface parallel 
 to the direction ${\bf e}_1$ or ${\bf h}_1$ is shown in Fig.\ref{unit_cells}a (horizontal line) or in Fig.\ref{unit_cells}b 
(vertical line), respectively. 
We found by direct calculation  $\sigma=2J_2/3$ or  $\sigma =J_2/{\sqrt 3}$ 
for the interface parallel to ${\bf e}_i$ or  to ${\bf h}_i$, respectively.
Thus, the most favorable are the lines parallel to the lattice vectors  ${\bf h}_j$.

When $c$ is slightly larger from zero or slightly smaller than  $1/3$, then a droplet in the vacuum or a vacancy 
in the close-packed system is created, with the shape
determined by the minimum of $H_s=\sum_i \sigma_iL_i+ \sum_jV_j$ under the constraint of fixed area of the droplet or the vacancy.
In the above expression, $\sigma_i$  and $L_i$ are the surface tension and 
the length of the segments with the orientation $i$, respectively, and $V_j$ is the energy of the $j$-th vertex.
When the number of the particles is properly adjusted, the interface line has a hexagonal shape,
with the edges parallel to ${\bf h}_j$.
The surface energy per the perimeter $P=6k\sqrt 3$ of a droplet or a vacancy  is $H_s/P=(1+(2k)^{-1})J_2/\sqrt 3$
or  $H_s/P=(1-(2k)^{-1})J_2/\sqrt 3$, respectively. 
We can see that the  hexagonal void is more-
and the hexagonal droplet is less favourable than the strait line interface, especially at small $k$. 
In Fig.\ref{unit_cells}c, a part of the snapshot obtained in MC simulations of the system with $c=395/1296$ at $T=0.1$ is shown. 
One can see that the sides of the  hexagonal void are parallel to the lattice vectors ${\bf h}_i$.   

Let us focus on the interface between the hexagonal phase of particles (c=1/3) and the hexagonal phase of voids (c=2/3). 
The interfaces parallel and perpendicular to the lattice vector ${\bf e}_1$ are shown in Fig.\ref{interphase}. 
The surface tension for the orientation of 
the interface  parallel to the lattice vector ${\bf e}_1$ can be easily calculated, and is $\sigma =2J_2/3$. 
 For the lines of the second type, the structure of the left  interface is different from the structure of the right one 
 (Fig.\ref{interphase}b). 
 When the number of the lattice sites in each row is decreased by one and the column to 
 the left of the left interface is removed, 
 the left interface will have the same structure as the right one, up to the mirror symmetry. 
 Similarly, by keeping the left interface unchanged, and inserting a mirror image of the previously removed column to
 the right from the right interface,  we obtain two identical interfaces. Using (\ref{sig}) we calculated $\sigma$ for 
 the geometry shown in Fig.\ref{interphase}b as well as for the other two geometries mentioned above, and obtained
 in each case the same result as for the interface between the  vacuum and $c=1/3$ phases, namely $\sigma_2=J_2/\sqrt 3$.
The Monte Carlo (MC) simulations for low temperature ($T=0.1$) confirm the preference of the interface lines parallel 
to ${\bf h}_i$ (Fig.\ref{fig:sublattices}).


\section{The GS of model II (thick shell)}
\label{GSII}

\subsection{The GS of an open system}
\label{GSIIa}
\begin{figure}
	\centering
	\includegraphics[scale=0.5]{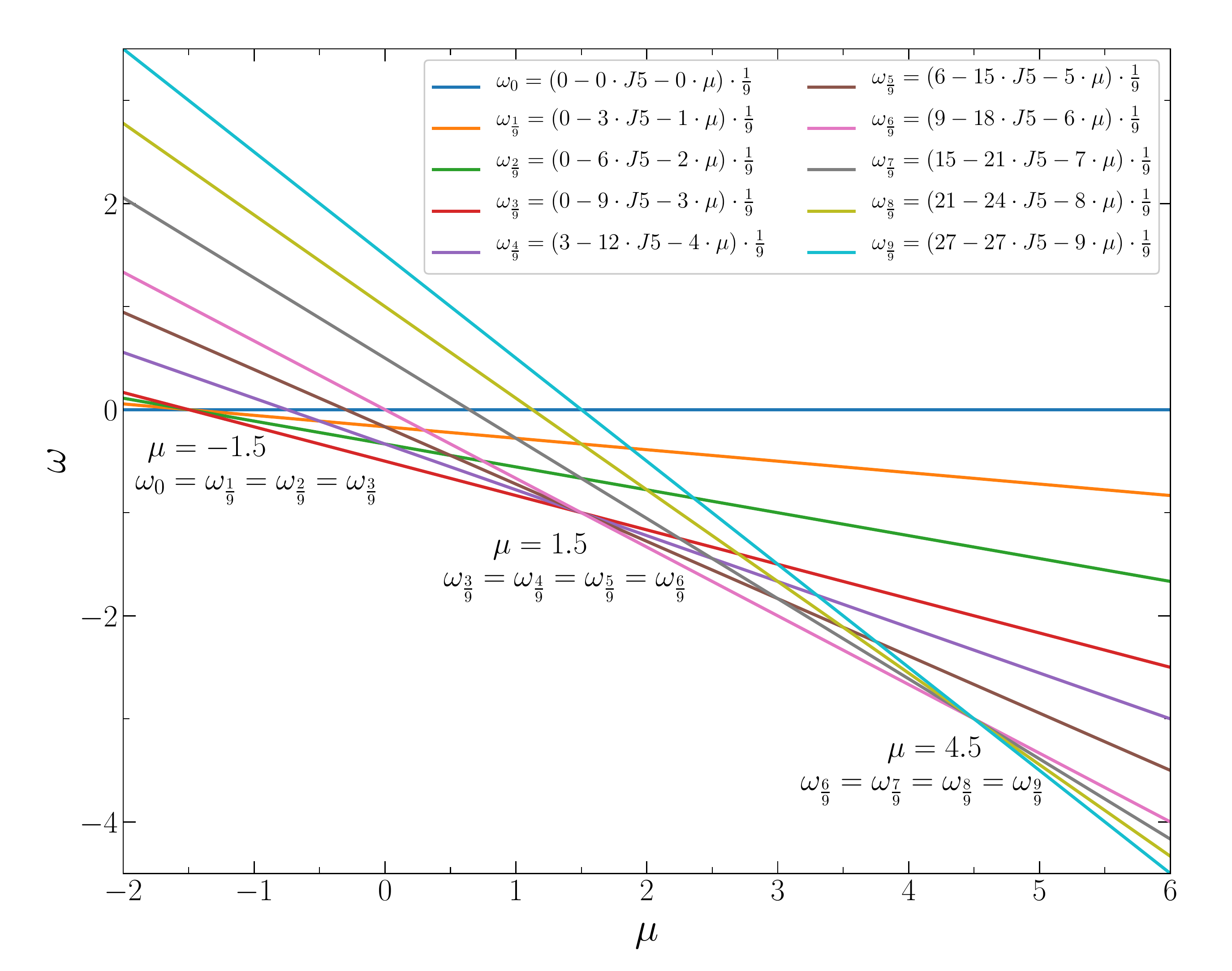}
	\caption{The dimensionless grand potential per lattice site, $\omega$,  for $J_5=1/2$ and 
	different concentrations, as a function of dimensionless
	$\mu$.	Four phases can coexist at the dimensionless chemical potential values $-3/2, 3/2, 9/2$.
	The structure of the phases is shown in the cartoons in Fig.\ref{fig:GS}.}
	\label{fig:energies}
\end{figure}

For the system with the repulsion of the first neighbors and attraction of the fifth neighbors, the unit cell contains
$3\times 3$ lattice sites and each of its nine sites has six nearest images in the nearest unit cells (see Fig.\ref{fig:sublattices}b).
Thus, these nine 
sites generate nine sublattices and the successive filling of the sublattices generates concentrations $0, 1/9, 2/9,...9/9$. 
The corresponding ordered patterns are shown in Fig.\ref{fig:GS}.
We can see the hexagonal phases with the lattice constant equal to the diameter of the hard-core, of the inner-, or of the outer
 shell for $c=1$, $c=1/3$ or  $c=1/9$, respectively, and
the honeycomb lattice or the lattice of rough clusters for $c=2/9$ or $c=4/9$. The same
structures, but with empty sites replacing the occupied ones and vice versa are also present, giving together 10 possible phases. 
Comparison of $\omega(c)$ for $c=n/9$ shows that for any fixed value of  $J_a$ only the concentrations $0, 1/3, 2/3$ and $1$
can be realized
for some range of $\mu$. The coexistence
lines on the $(J_a,\mu)$ phase diagram are the same as in the previous case, but with  $J_a=J_5$ replacing $J_a=J_2$. 
However, at the coexistence lines between the phases with $c=n/3$ and $(n+1)/3$ with $0\le n\le 2$, two more phases,
with $c=(3n+1)/9$ and $c=(3n+2)/9$ are stable too (Fig.\ref{fig:GS}). 
For fixed $J_5$, these phases are stable for a single value of $\mu$ only,
and coexist with the other two phases. The dependence of $\omega(c)$ on $\mu$ for ten values of $c$ and 
for $J_5=0.5$ is shown in Fig.\ref{fig:energies}.
One can see that  for three values of $\mu$, namely $\mu=-3/2, 3/2, 9/2$,
four out of ten lines intersect, and that the value of $\omega(c)$ for 
the remaining six values of $c$ is larger. Similar behavior is found for all values of $J_5$. 
Thus, four phases coexist at the coexistence lines shown in Fig.\ref{fig:GS}. 



Note that in the absence of interactions between the second neighbors,  $\omega(c)=-c(\mu +3J_5)$ for $c\le 1/3$.
This is because  in the considered structures (see  $c\le 1/3$ in Fig.\ref{fig:GS}) 
the particles occupying one sublattice do not interact with the particles occupying another one. 
For $\mu +3J_5=0$,  $\omega(c)=0$ 
for all the phases with $c\le 1/3$. This degeneracy could be removed if the interaction between the second neighbors 
were present. For the other two cases of four-phase coexistence, the degeneracy has similar origin.

\subsection{The GS for fixed number of particles and the surface tensions}
 \label{STII}
 
 In an open system, the GS $(\mu,J_a)$ of model I and II differ only at the coexistence lines that are  regions of zero-measure. For fixed number of particles, however, the stable structures can be completely different in the two models. In particular, for  $N=M/9$ an interface between vacuum and the 
 $c=1/3$ phase is formed in model I, 
 whereas in model II the phase with $c=1/9$ occupies the whole lattice.
 Similarly, for  $N=kM/9$ with $k=2,4,5,7,8$,
  two-phase coexistence with an interface occurs  in model I (see Fig.\ref{interphase}b for $k=4,5$), but in model II, 
  the  periodic 
  phase with $c=2/9,4/9,5/9,7/9,8/9$, respectively, is present (Fig.\ref{fig:GS}, bottom row).

For $N\neq kM/9$ with $k=0,1,...,9$, the phase coexistence of the two phases, the concentrations of which are the closest 
to the mean concentration, occurs in model II. 
At the overall concentration $0<c<1/9$, the phase with the concentration $c=1/9$ coexists with the vacuum phase. 
The line tension is $\sigma=J_a/3$ and $J_a/\sqrt 3$ for the lines parallel to the vectors  ${\bf e}_i$ and  ${\bf h}_i$, respectively. 
Thus, the interface parallel to ${\bf e}_i$  is more preferable. 
Interestingly, in both models
the orientation of the interface is determined by the hexagonal lattice formed by the particle cores, regardless 
of its orientation with respect to the underlying triangular lattice, and the particles at the boundary lie on a straight line.
   
  When $c$ is close to 0, the particles can form  rhomboidal or hexagonal clusters, in both cases with sides parallel to 
  the unit lattice vectors ${\bf e}_i$. 
  By comparing the surface energies for polygons of (approximately) the same area, we have found that 
 the hexagonal clusters are more preferable.
 
  At a fixed concentration slightly below $1/9$, hexagonal or rhomboidal voids
  can be created. We have found that a small rhomboidal void  is more preferable than a hexagonal one.
  %
%
 The energy difference is rather small, however ($-J_5$); rhombuses and irregular hexagons 
 with the sides parallel to the lattice vectors ${\bf e}_i$  correspond to local minima of the energy, and may appear 
 in simulations at low temperature. For large empty spaces, the hexagonal configuration is again more preferable.

In Ref.\cite{groda:19:0} the model of intermediate shell thickness was considered, with attraction of the third neighbors 
instead of the second (model I) or the fifth (model II) ones. The unit cell contained four lattice sites, and 
five phases with concentrations 0, 1/4, 1/2, 3/4, 1 were present in the GS.
The symmetry of this intermediate model is 
similar to the symmetry of model II; the vectors connecting the third neighbors are parallel to the lattice vectors ${\bf e}_i$. 
All the results and conclusions concerning the interface lines and preferable configurations for model II remain 
valid for the intermediate model. In calculating the line tensions, we need to take into account the distance $2a$ 
between the third neighbors and the line tensions of model II have to be multiplied by 3/2.       

\section{The thermodynamics of the system for $T>0$}
\label{T>0}
At low dimensionless temperatures $T=k_BT^*/J$ (where $T^*$ is the absolute temperature and $k_B$ the Boltzmann constant), 
the 
ordered states depending on the chemical potential or density remain present.
In this section we present the $\mu(c)$ isotherms, isothermal compressibility, specific heat and order parameter
 obtained for both models by MC simulations for $J_a=0.5$ and a range of $T$. The Metropolis importance sampling 
simulations were performed for the system of $96\times 96$ lattice sites with periodic boundary conditions. 
1000 Monte Carlo simulation steps (MCS) were used for equilibration. The subsequent 10 000 MCS were used 
for calculating the average values.

\subsection{Model I (thin shells)}
\begin{figure}
        \centering
        \includegraphics[scale=0.8]{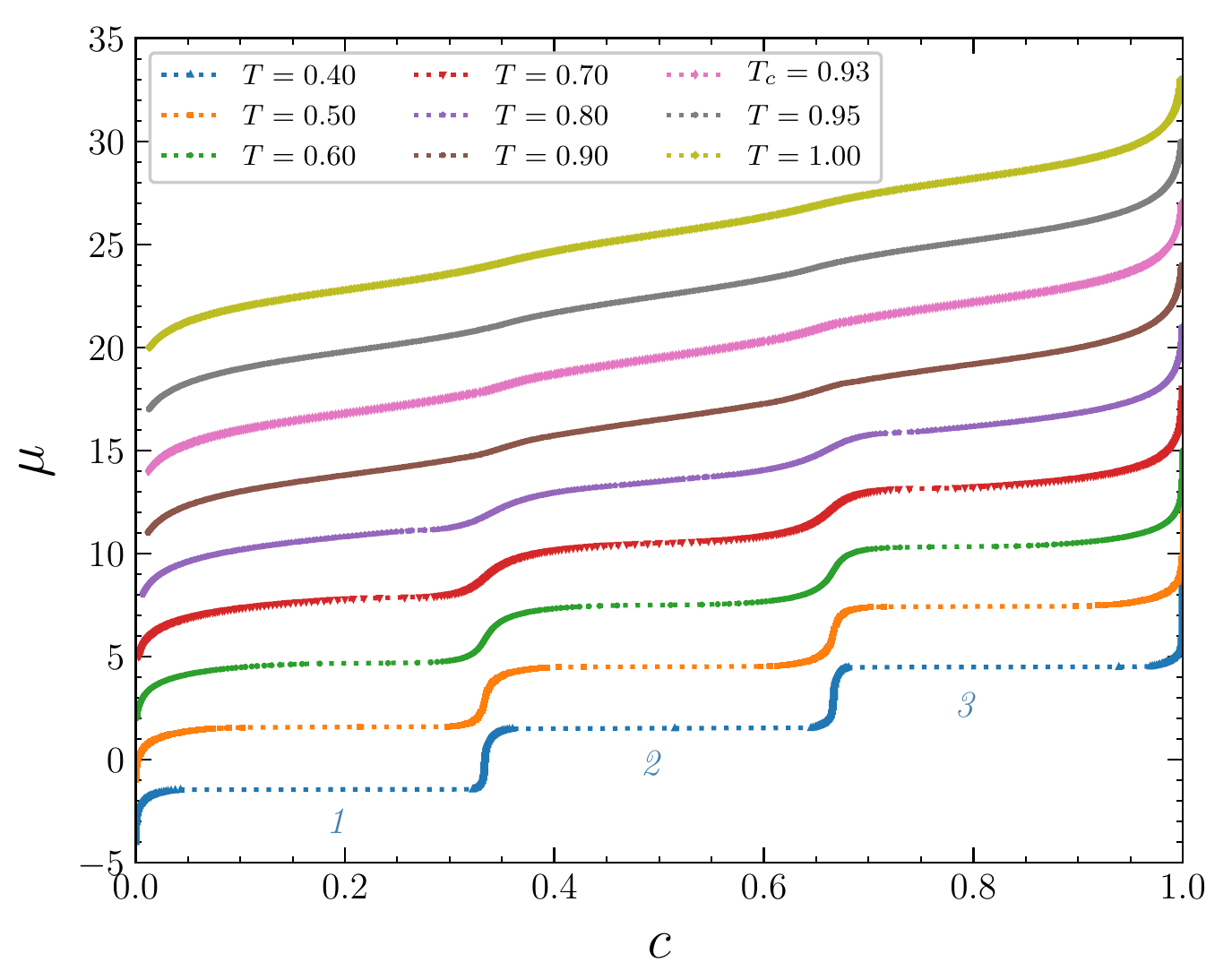}
        \caption{The chemical potential - concentration isotherms for model I at $J_2=0.5$. 
        The isotherms are shifted in the vertical direction by 3 
        from each other for clarity. The isotherm at $T=0.4$ is not shifted.}
        \label{isotherms}
    \end{figure}

\begin{figure}
        \centering
        \includegraphics[scale=0.5]{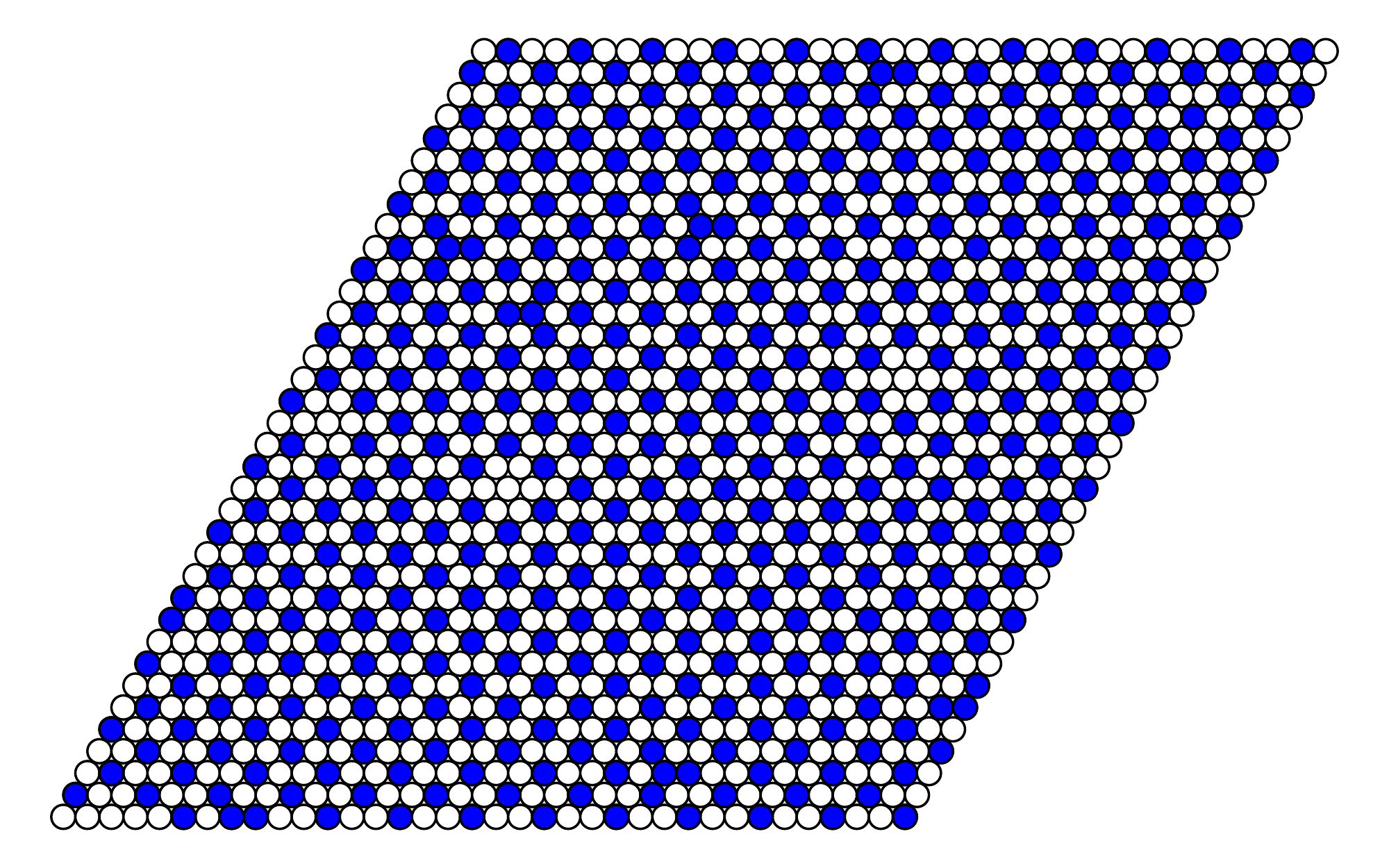}
        \caption{The snapshot of model I for $\mu=0$ and $T=0.6$, after 1000 MCS for equilibration and 2000 simulation MCS.
        $c=435/1296$ and $\eta\approx 0.98.$ There are  427 occupied and 5 vacant sites 
        on the main sublattice, and 8 occupied sites on the other two sublattices.}
        \label{fig:defects} 
\end{figure}

The concentration isotherms for model I are shown in Fig.\ref{isotherms}.
On the isotherms for $T\leq 0.7$ it is clearly seen that there are no simulation points on large intervals of the concentration. 
These dotted horizontal lines correspond to a coexistence of two ordered phases, namely vacuum and $c=1/3$, next $c=1/3$ and $c=2/3$, 
and finally, $c=2/3$ and dense. A few points in these intervals correspond to metastable states that occasionally can be realized
in the course of simulation. The concentration intervals corresponding to the stable phases increase with temperature
due to thermally induced structural defects. Some points at the ends of the stable phases can also correspond to metastable 
(superheated or supercooled) states.  
 Fig.\ref{fig:defects} demonstrates the defects in the $c=1/3$ phase in model I at fixed $\mu=0$ and $ T=0.6$, with
 the average concentration close to $1/3$.
 One sublattice is completely occupied in the GS at this value of the chemical potential. 
 Thermal fluctuations result in a few voids on the main sublattice and a few occupied sites on the other ones.
 The number of defects increases with increasing temperature and/or variation of the chemical potential.   
   
\begin{figure}
        \centering
        \includegraphics[scale=0.8]{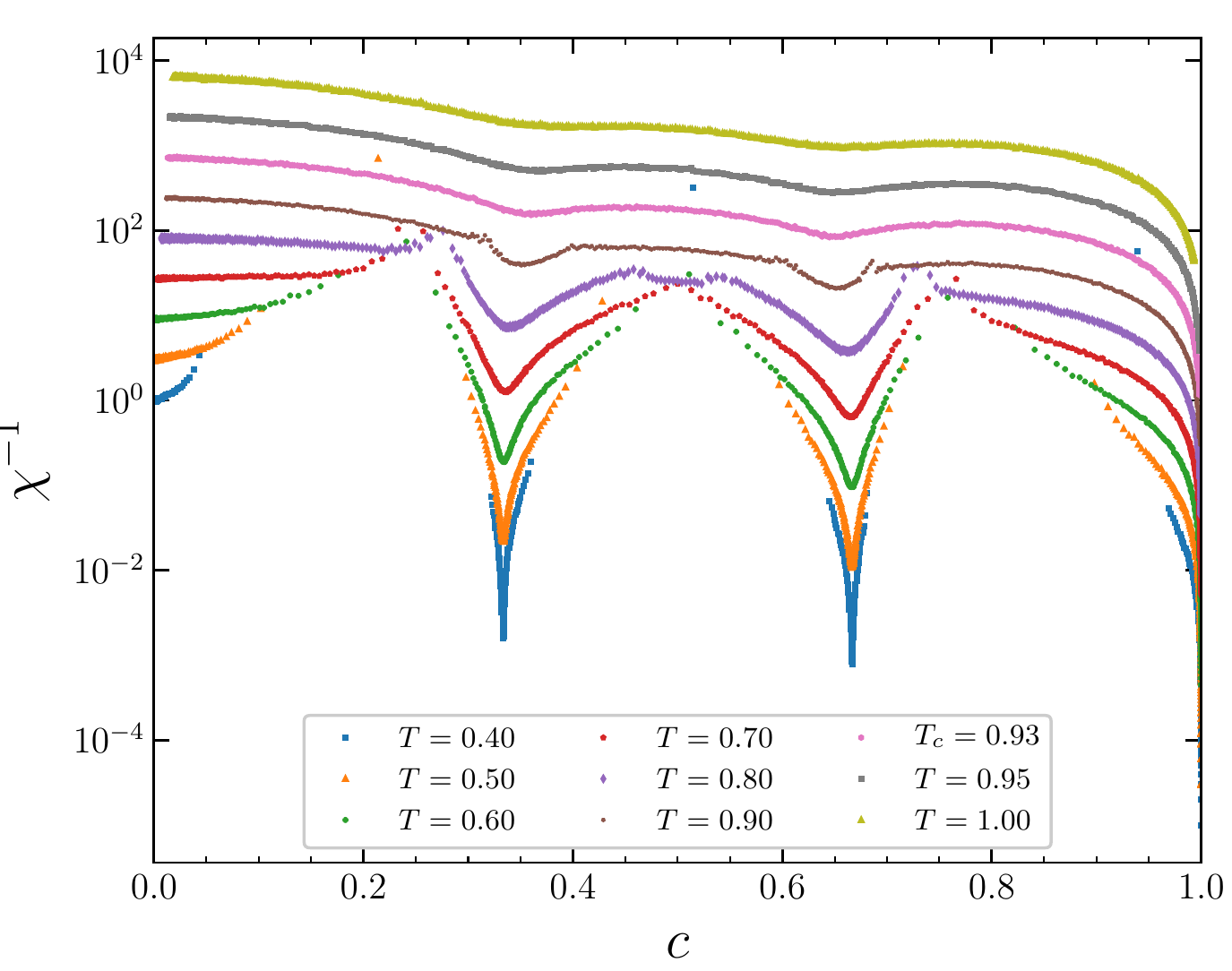}
        \caption{The inverse thermodynamic factor $versus$ concentration for model I at $J_2=0.5$. 
        The curves are shifted in the vertical direction by $3^n$ from each other for clarity. The curve at $T=0.4$ is unshifted.}
        \label{cfluct}
\end{figure}

At low temperatures, small variations of the concentration at large variation of the chemical potential are observed in
the ordered phases, signaling large values of the thermodynamic factor $\chi_T=c(\partial(\beta \mu)/\partial c)_T$. The  thermodynamic factor is inversely proportional to the isothermal compressibility $\kappa_T=(\partial c/\partial p)_T/c$ (where $p$ is pressure) that in turn is proportional to the
concentration fluctuations,
 \begin{equation}
    \label{chi}
\frac{\langle (N- \langle N \rangle )^2 \rangle }{\langle N \rangle}  =\chi_T^{-1}=Tc\kappa_T.
 \end{equation}
In the above, the angular brackets $\langle ... \rangle$ mean the ensemble or MC simulation average. 
Thus, the concentration fluctuations are suppressed in the most ordered states and can reach large values 
at the phase boundaries (Fig.\ref{cfluct}).

\begin{figure}
        \centering
        \includegraphics[scale=0.8]{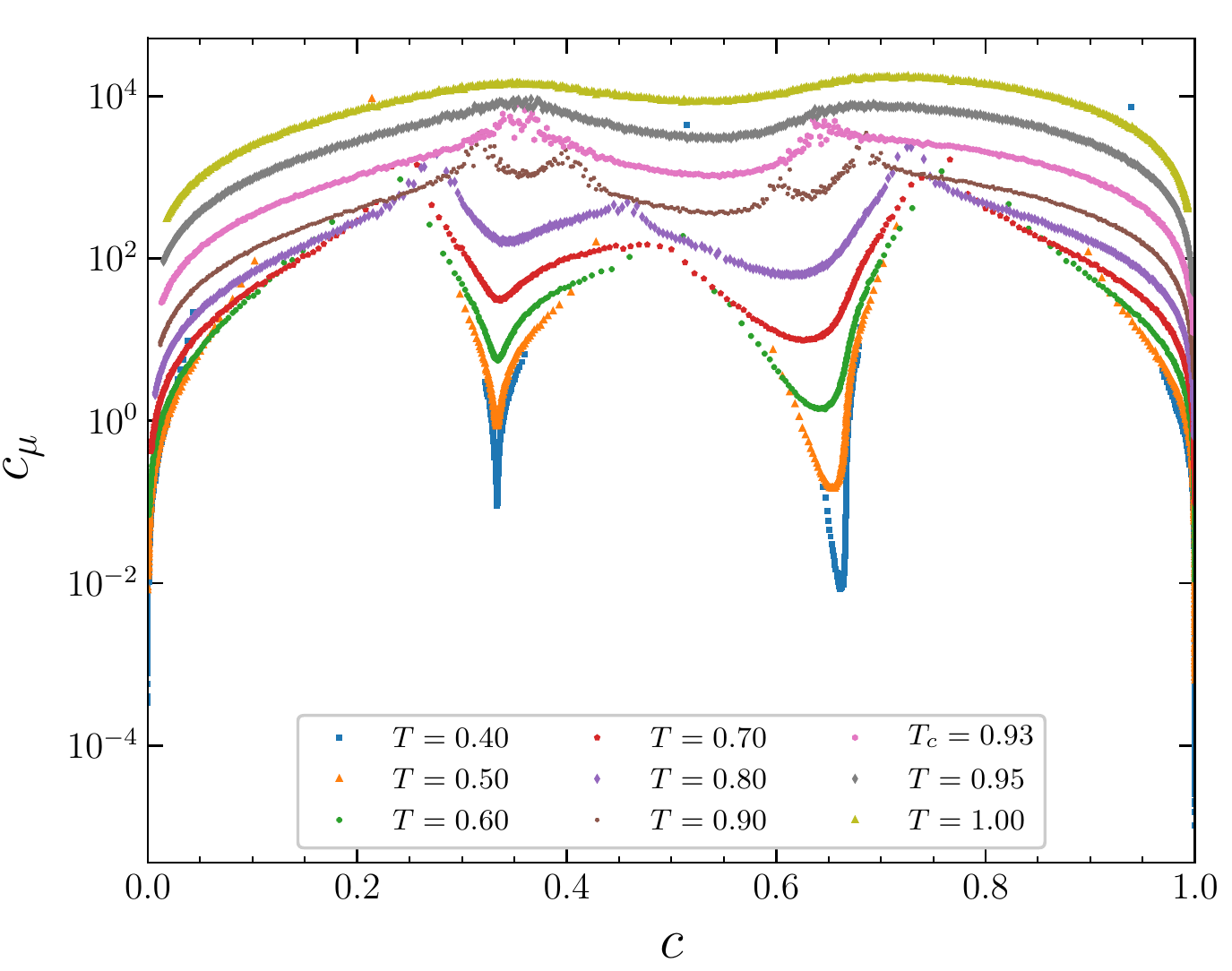}
        \caption{The concentration dependence of the specific heat  for model I at $J_2=0.5$. 
        The curves are shifted in the vertical direction by $3^n$ from each other for clarity. The curve at $T=0.4$ is unshifted.}
        \label{efluct}
\end{figure}

The maxima of the compressibility (minima of the thermodynamic factor) can serve as an indicator of the phase transitions 
in the finite system. 
Another indicator of the phase transitions is the energy fluctuation or the specific heat 
that in the dimensionless form can be written as

 \begin{equation}
    \label{heatcap}
c_\mu=\frac{ C^*_\mu }{ k_B\langle N \rangle }=\frac{1}{k_B\langle N \rangle} \Bigg(\frac{\partial E^*}{\partial T^*}\Bigg)_{\mu} =\frac{\langle (E- \langle E \rangle )^2 \rangle  }{\langle N \rangle T^2},
\end{equation}
where $C^*_\mu$ is the heat capacity with constant chemical potential, and $E=E^*/J$ is the dimensionless system energy.
Again, the energy fluctuations are suppressed in the most ordered states, and are large at the phase transition points (Fig.\ref{efluct}).
The concentration dependence of the heat capacity is qualitatively changed at the temperature 0.93 when the minima disappear
at concentrations close to 1/3 and 2/3. Thus, $T_c=0.93$ can be considered as the critical temperature.
This conclusion is supported by the behavior of the concentration isotherms (Fig.\ref{isotherms}) and concentration
fluctuations (Fig.\ref{cfluct}).

The ordered states of the system 
are characterized by the order parameter (OP) that can be calculated as 

\begin{equation}
    \label{orderparameter1}
\eta=3\langle N_1 \rangle/M - 3 \langle (N_2 + N_3) \rangle /2M \quad \rm {at} \quad c<1/2
\end{equation}
or
\begin{equation}
  \label{orderparameter2}
\eta=3\langle (N_1+N_2) \rangle/2M - 3 \langle N_3 \rangle /M 
\quad \rm{at} \quad c>1/2,
\end{equation}
where $N_1 \ge N_2 \ge N_3$ are the numbers of the particles on the sublattices. 
 The OP is close to 1 if one or two sublattices are almost completely occupied,
 while  the remaining two or one, respectively, 
 are almost empty. If the sublattices are almost equally occupied, the OP is close to zero.

\begin{figure}
        \centering
        \includegraphics[scale=0.8]{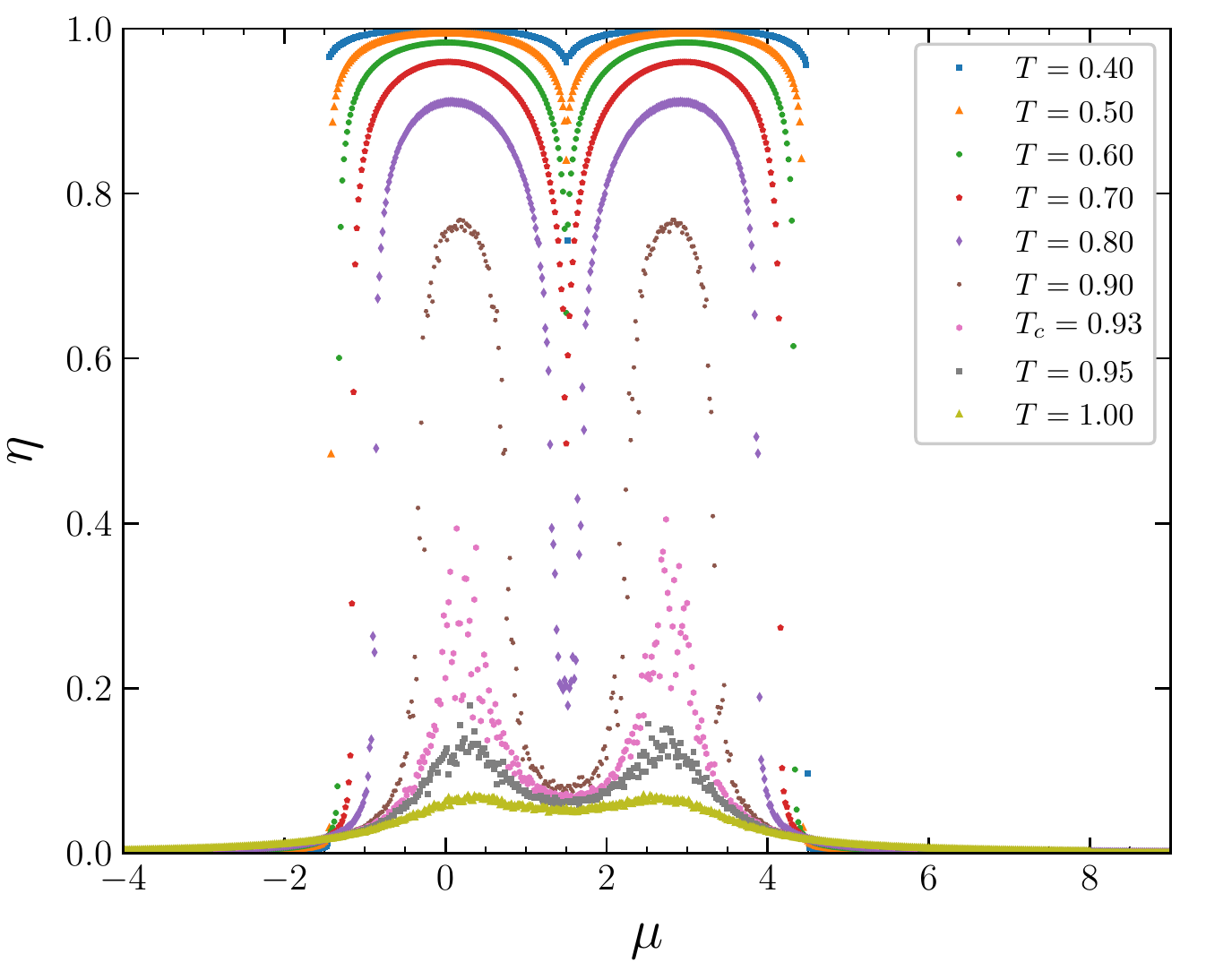}
        \caption{The order parameter (Eqs.(\ref{orderparameter1})-(\ref{orderparameter2}))
        for  model I at $J_2=0.5$ $versus$ chemical potential. 
      }
        \label{orderpar}
\end{figure}

The most ordered structure occurs for  $\mu\approx 0$ or $\mu\approx 3$, i.e. in the center of the stability region of the
$c=1/3$ or $c=2/3$ phases 
(Fig.\ref{orderpar}).
The OP decreases for increasing $T$ and/or for $\mu$ departing from $\mu= 0$ or $\mu= 3$, and experiences
large fluctuations
 at the critical isotherm ($T_c\approx 0.93$) for the average concentration $c=1/3$ or 2/3. 
 The OP at the temperatures above the critical one remains different from zero. It is  partially
due to the calculation procedure, and partially due to a more complicated phase behavior of the model in this temperature
range \cite{mihurat:77:0, landau:83:0}. As we are interested in the ordered patterns and the model is valid for a limited range of $T$,
we do not study the high-$T$ properties in more detail. 

\subsection{Model II (thick shells)}

\begin{figure}
        \centering
        \includegraphics[scale=0.8]{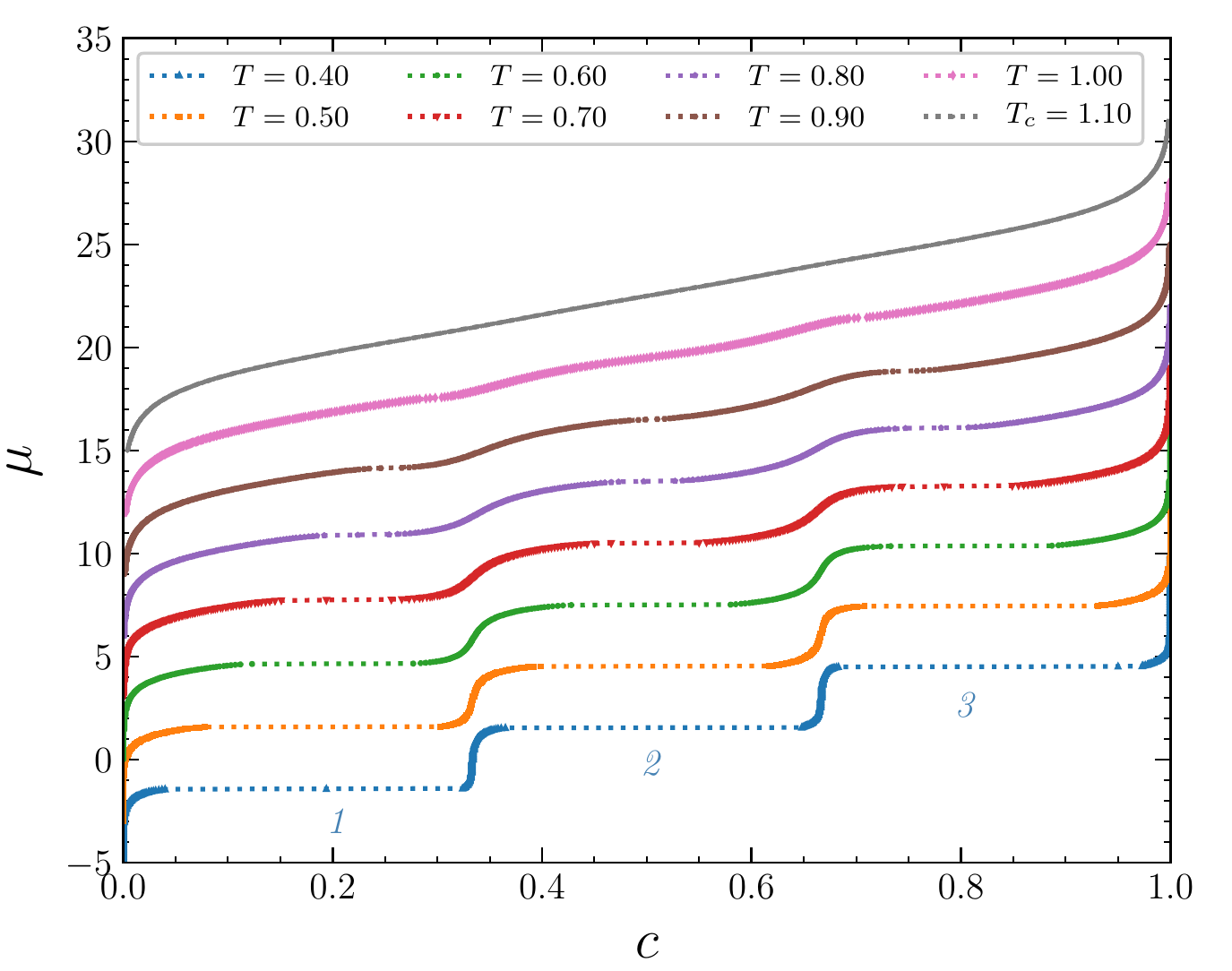}
        \caption{The concentration isotherms for  model II at $J_5=0.5$. The isotherms are shifted in the vertical 
        direction by 3 from each other for clarity. The isotherm at $T=0.4$ is not shifted.}
        \label{fig:isotherms1-5}
\end{figure}

The concentration isotherms for model II (Fig.\ref{fig:isotherms1-5}) look very similar to that for model I, although the critical 
temperature is a little bit higher ($T_c \approx 1.10$). This is the result of more space for the cores of the particles in model II, where 
 the second, third and fourth neighbors can be occupied at no energetic cost. 
 No indication of ordered phases except from $c$=1/3, 2/3 or 1 is seen in these isotherms. 

\begin{figure}
        \centering
        \includegraphics[scale=0.8]{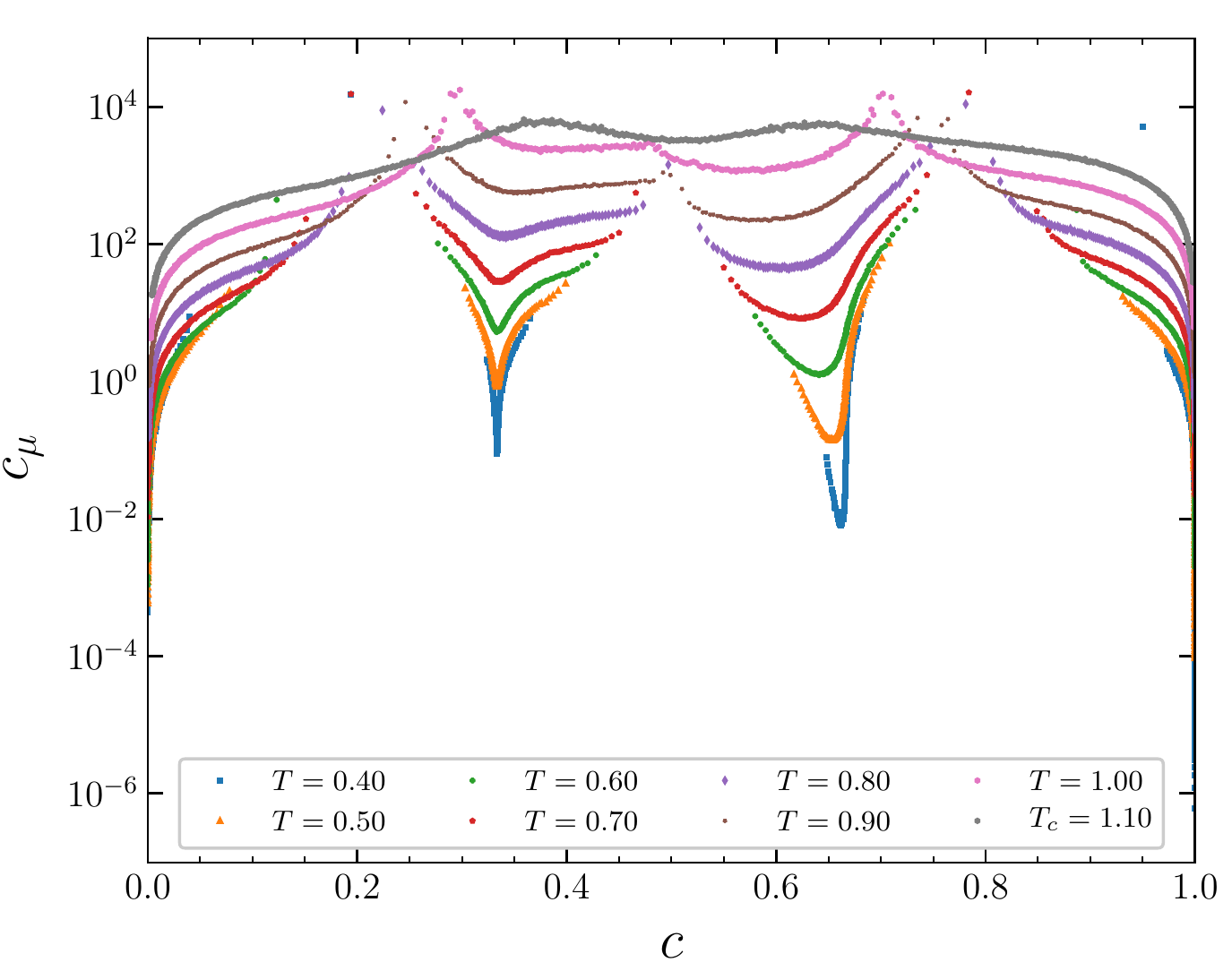}
        \caption{The concentration dependence of the heat capacity for model II at $J_5=0.5$. The curves are shifted 
        in the vertical direction by $3^n$ from each other for clarity. The curve at $T=0.4$ is not shifted.}
        \label{fig:heatcap1-5}
\end{figure}

 Again, the heat capacity has minima at the concentrations 0, 1/3, 2/3 and 1 (Fig.\ref{fig:heatcap1-5})
 that confirms the appearance in the open system of these phases only.
 The same conclusion follows from the concentration fluctuation isotherms and OP behavior.
 The character of fluctuations in both systems is similar. 
 The other ordered phases are hidden in the horizontal segments of the $\mu(c)$ isotherms.
Each dotted line segment in Fig.\ref{fig:isotherms1-5} represents several phase transitions.
The first segment represents the sequence:
disordered dilute gas phase $\rightarrow$ $c=1/9$ 
phase $\rightarrow$ $c=2/9$ phase $\rightarrow$ $c=1/3$ phase.  The second segment represents the sequence: $c=1/3$ phase
$\rightarrow$ $c=4/9$ phase  $\rightarrow$ $c=5/9$ phase $\rightarrow$ $c=2/3$ phase. Finally, the third 
segment corresponds to the sequence: $c=2/3$ phase $\rightarrow$ $c=7/9$ phase $\rightarrow$ $c=8/9$
phase $\rightarrow$ condensed phase with voids. Thus, four coexisting phases at particular values of 
the chemical potential in the ground state evolve into succession of phase transitions while the lattice 
concentration is growing at $T>0$.  For different $\mu$ values,
$\omega$ takes local minima for the phases absent in model I. 

When the number of particles $N$ is fixed at  low $T$, either
a particular phase exists or two phases with the densities  closest to $N/M$ coexist and an interface between them occurs, as in the GS. 
In 2D systems there exists a long-wavelength interface instability due to capillary 
waves \cite{buff:65:0, zittarg:67:0, widom:72:0, binder:83:0}. 
This instability leads to rich variety of different patterns when $T$ increases and $N$ is fixed, 
because many ordered patterns are metastable in model II. For example,
at $c=1/6$, $T=0.5$ and $N$ fixed, 
large dynamical fluctuations 
in the system exist along the MC simulation trajectory.
In the sea of the $c=1/9$ phase, islands of the phases $c=2/9$, $c=1/3$ and even vacuum in different transient configurations may occur.    

\section{Discussion and conclusions}

The purpose of our study was determination of the effect of the shell thickness and softness on patterns formed by core-shell
particles adsorbed at an interface between two liquids. We assumed that  when the shells of the two particles touch each other,
an effective attraction between the particles is 
induced by capillary forces, as  suggested by the results of Ref.\cite{rauh:16:0}.
 We compared patterns formed in model I, 
where the hard-core of the particle is covered by a relatively stiff shell, with patterns formed in model II, where
the above mentioned inner shell is surrounded by a much softer outer shell, and the whole particle is bigger. 
The shell thickness is temperature-independent in our models, and the results can be valid for a limited range of $T$, so that
no transition in the shell structure takes place.

The phase diagrams of both models in the temperature -- chemical potential variables are very similar. At low $T$,
dilute gas or closely packed cores  appear for very small or very large values of the chemical potential, respectively.
For intermediate values of $\mu$, 
a hexagonal lattice of particles or holes, 
with the lattice constant equal to the diameter 
of the inner shell, and with concentrations $c\approx 1/3$ or $c\approx 2/3$, respectively, can be formed. 
Only at three  values of 
the chemical potential, corresponding to phase coexistences (the horizontal segments of the $\mu(c)$ isotherms in 
Figs. \ref{isotherms} and \ref{fig:isotherms1-5}),
the properties of the two models are different. The transition between the $ c=0$ and $ c=1/3$ phases in model I is replaced
by the sequence of the transitions $c=0\to c=1/9$, next $c=1/9\to c=2/9$, and $c=2/9\to c=1/3$ in model II. In the 
$c=1/9$ and $c=2/9$ phases, the cores occupy sites of the hexagonal lattice
with the lattice constant equal to the diameter of the outer shell and the  honeycomb lattice, respectively
(see Fig.\ref{fig:GS} for the structure of the phase referred to as ``$c=n/9$ phase''). 
All these transitions occur at the same value of $\mu$,
and are hidden in the horizontal segments in Fig.\ref{fig:isotherms1-5}. 
Similarly, the other two horizontal segments in 
Figs. \ref{isotherms} and  \ref{fig:isotherms1-5} correspond to the transition between two phases in model I, and to a sequence of 
3 phase transitions in model II.

The hexagonal lattices with different lattice constants are present in many experimental systems
~\cite{isa:17:0,honold:15:0,rauh:16:0,vogel:12:0,karg:16:0,geisel:15:0,nazli:13:0,volk:15:0}.
The honeycomb lattice was obtained experimentaly
by sequential deposition of Au-Au and Ag-Au PNIPAM particle monolayers~\cite{honold:16:0}, 
and the clusters were seen in Ref.\cite{rauh:16:0}. 

From our results it follows that
the shape of the isotherm can be very misleading, when particular patterns can occur for  very small intervals of $\mu$, and the steps
on the isotherms are hardly visible.
Even when the additional phases are present for very small intervals of $\mu$, or as in model II for single values of $\mu$,
they strongly influence the structure for fixed number of particles. 

When at low $T$ the number of particles is fixed, the patterns formed in models I and II 
are the same
only for the values of the concentration $c$ corresponding to stability of the  phases of model I.
For large intervals of $c$, patterns formed in models I and II are different.
In particular, for the whole interval $0<c<1/3$, an interface between the dilute gas and the $c=1/3$ phase occurs in model I.
In model II, the interface is formed  between the dilute gas and the $c=1/9$ phase for  $0<c<1/9$,
next, between the  $c=1/9$ and  $c=2/9$
phases for $ 1/9<c<2/9$, and finally between the   $c=2/9$ and  $c=1/3$ phases for $ 2/9<c<1/3$. 

The surface tension between  the ordered phases depends on the orientation of the interface. 
We have found that the surface tension takes the smallest value
for the  interface parallel to the side of the hexagon in all the hexagonal phases in both models 
and in the model of Ref.\cite{groda:19:0}.
For this orientation, the boundary particles
lie on a straight line,
in agreement with Ref.\cite{rauh:16:0}.
This result is independent of the orientation of the 
hexagonal structure with respect to the underlying lattice.

Our results show that particles with composite shells consisting of a stiff  inner shell and a soft outer shell can
form  complex patterns, including the honeycomb lattice
and periodically distributed clusters, in addition to the common hexagonal lattices with smaller and larger unit cells. 
 Modified model II, with the second-, third- and fourth neighbor 
 interactions representing modifications of the structure of the composite shell of the particles,
could lead to an expansion of  the stability region of the additional phases from a single value to  an interval of $\mu$.
The horizontal segment in Fig. \ref{fig:isotherms1-5} could evolve into 3 segments separated by steps.
The question of the sensitivity of the shape of the isotherms to the effective interactions following from the structure of 
the composite shell requires further studies. This is the goal of our future work.

\section{Acknowledgements}
 This project has received funding from the European Union Horizon 2020 research 
and innovation programme under the Marie
Sk\l{}odowska-Curie grant agreement No 734276 (CONIN).
An additional support in the years 2017-2020  has been granted  for the CONIN project by the Polish Ministry
of Science and Higher Education.

\end{document}